\documentclass[12pt]{article}

\begin{document}

\begin{center}

{\bf \Large Search for Life on Exoplanets: \\
Toward an International Institutional Coordination\footnote{Presented at the  Workshop {\it Exoplanets and Disks: Their Formation and Diversity}
Kailua-Kona, Hawaii, March 2009}}\\
\end{center}

\bigskip 

\begin{center}
{\it \large Jean Schneider\footnote{Jean Schneider@obspm.fr}, Vincent Coud\'e du Foresto, Marc Ollivier}\\
\end{center}

\bigskip

{\bf Abstract}

Searching for life in the universe will make use of several large space missions in the visible and thermal infrared, each with increasing spectral and angular resolution. They will require long-term planning over the coming decades. We present the necessity for building an international structure to coordinate activities for the next several decades and sketch the possible structure and role of a dedicated international institution.

\section{The Role of Exoplanets in Science}

The question "Are we alone in the universe ?" has been continuously raised
by hundreds of authors for the past 23 centuries \cite{Crowe} \footnote{The question of why has this debate
been restricted to ''western'' cultures is addressed by \cite{Schneider}.}. 

This question goes far beyond "standard" astronomy. Indeed, it opens the possibility to better understand in the (far?) future (thanks to some future discovery of other forms of life, and possibly "intelligent" life) how life is connected to abiotic matter\footnote{Let us make here a philosophical remark: in some interpretations of measurement in quantum physics, the measurement process does not result from a system-apparatus ordinary interaction described by a Hamiltonian, it is a {\it sui generis} event; similarly, in spite of some successes of the chemical reductionism of biology, one should not definitely rule out
the possibility that life is not reducible to physics.}. 

Thus, there are good reasons for putting the investigation of exoplanets in a special position in astronomy. 
This special position has been recognized within (among other places)  the U.S.
National Aeronautics and Space Administration (NASA). For instance, Ed Weiler's declaration 
after the Kepler mission launch \cite{Weiler}: "This is a historical mission. It's not just a science mission". 
As another example, the FY 2010 NASA Budget Request \cite{NASA} states that we are at the "threshold of a voyage of unprecedented scope and ambition, promising insight into one of humankind's most timeless questions".

Exoplanets are not the only fascinating domain in astronomy. For instance, dark energy is another one,
opening for the first time the possibility of post-Einsteinian developments of astrophysics. But, there is a great difference between exoplanets and dark energy. There are no precise perspectives about what to do after the energy density $w$ and its dependence on redshift and direction in the sky are measured in the 
next few decades. One cannot anticipate whether other fields in fundamental astrophysics (like e.g. quantum black holes?) will emerge or not. For exoplanets we already know what to do in the next hundred years.
Indeed, several ground-based facilities and space missions will be necessary to investigate exoplanets in the visible reflected light and by their infrared thermal emission. This will require several generations of instruments with increasing spectral and high angular resolution (including multi-pixel cartography).
For an outlook of this long term future see \cite{Schneideretal}.

\section{The Need for World-wide Coordination}

Yet, because the field of exoplanet studies is relatively new, and remains multidisciplinary in nature, it does not have the institutional visibility that would match the general interest in the theme,
and it is not yet seen as a discipline in its own right. More often than not, at all
levels of the scientific community, exoplanet science and exobiology appear implicitly as a subset of other disciplines - planetary sciences, instrumentation, biology etc. As a consequence, support for related projects is often fragmented, and each piece has to compete in its own field with already wellestablished
programs.
The situation is particularly critical for space missions. The current trend (bias)
among space agencies (ESA and NASA alike) is to select missions which are both
low-risk and with an immediate return for science. This tends to favor "more-of-the-same"
projects, each of which is a more elaborate version of a previous mission. The
community behind those projects is already well-organized, the risks can be easily
assessed, little new technology is involved, and the science case is easier to put
forward.
It is clear that the goal of spectroscopic characterization of the atmospheres of
habitable planets, in the search of biomarkers, will ultimately require one or more very
ambitious and innovative missions. Those cannot meet the feasibility criteria as
currently established by the space agencies, and the science return of a more
affordable demonstrator (one that would retire the risk on the bigger mission) cannot
meet the agency standards. This paradox implies that the current framework of mission
selection by NASA, ESA and the like is not compatible with a roadmap toward the
detection of life on other worlds.

Besides, this goal should not be Europe's, or America's, or only one continent's affair, but
should belong to humanity as a whole. Therefore, it appears desirable that its pursuit be
delegated to a single, transnational institution that would be dedicated to this objective.

Note that, beyond the present authors, this philosophy has already been 
endorsed by the Blue Dot Initiative \cite{Coude}.
\section{Possible Scenarios}
There is, among the community of planet hunters, an increasing desire emerging for an institutional world-wide coordination to make these projects a reality. The exact form is not yet clear, but it must certainly go beyond simple inter-agency bilateral "Letters of Agreement". In other words, what we need beyond collaboration is coordination by, for instance, an international institution like IPCC (Intergovernmental Panel on Climate Change). The contours of this institution could be discussed and clarified in the near future.

An excellent occasion to make this happen is the coincidence of the International Year of Astronomy 2009 
and the Pathways Conference in September \cite{Pathway}. Support or encouragement from the International
Astronomical Union would also be welcome.

An Exoplanet Institute (or whatever its name) would be both a scientific center and a
policy-making organization. It would carry enough weight to become the natural partner of
the different ground and space agencies in order to build collaborations between
them and coordinate their scientific and technical efforts. It should
receive enough support from its member states to be a structuring force in the field
-- some of that funding could actually go back to the respective agencies in
exchange for their participation, which would ensure the authority of the Institute
over those matters.
The first step towards the creation of the Exoplanet Institute is the clear expression of
such a need, stated in a solemn declaration by a group of world-class scientists.

Now, designing extensive plans for such a coordination is certainly premature.
To show that the concept is not empty, one can nevertheless sketch one or two scenarios.
For instance, one could take the example of the "Mandatory Science Program" of the European Space Agency. There, the countries participating in ESA commit themselves to fund the science projects decided by the
ESA Executive with the help of advisory bodies chosen among members of the astronomy community.
In return, member state industries have the garantee of contracts forwork, with amounts proportional
to member states' financial contribution to ESA. As an illustration, such a mechanism could be extended at the world level. It would not require extra budgets, just the transfer of existing budgets to the
coordinating entity. With a global world budget of a few hundred million Euros per year,
coming from contributions from all continents, one can launch one big mission every ten
years plus one or two medium mission every five years.
Another, much "softer"  possibility could be a permanent Interagency Group devoted to exoplanets, like the existing Inter-Agency Consultative Group (IACG). It would perhaps be less efficient than a "mandatory program", but it would constitute a starting point that could lead to the preparation of more elaborate scenarios.

Whatever the details of these suggestions, one may expect some {\it a priori} reservations. First, one may fear that astronomers may view this as a attempt to privilege exoplanet science compared
to other fields of astronomy, with the consequence that budgets in other fields may be decreased to fund this program. But, as explained above, there would not be extra budgetary outlays in each country or at participating institutions, just a world-wide coordination with a long-term view and standard budgets.
Second, some may feel that creating a new entity would lead to more bureaucracy. The administrative logistics would be not heavier than those of the IACG.
One may also object that we need to have at least some provisional agreement by the various national space agencies before trying to proceed with such a novel approach. That is true and that is why 
it is necessary to approach the agencies to inform them about this philosophy.

\section{Conclusion}
The purpose of this effort is to unite the global community of planet hunters, to have them join together in pursuit of a common goal. Here we have made only some illustrative suggestions that are open for debate.
To the readers who approve the spirit of this approach, we make a "call for support." Such support can be made by various means, such as papers in appropriate places or expressions of support toward the appropriate agencies and national institutions. Any comments to the present authors \footnote{Vincent.Foresto@obspm.fr, Marc.Ollivier@ias.fr, Jean Schneider@obspm.fr} are welcome.


\begin{thebibliography}{9}

\bibitem{Coude}V. Coud\'e du Foresto. Initial Report from the European Blue Dot team. (2009). in 
{\it Missions for Exoplanets: 2010-2020}. Pasadena, USA. (2009)
Availabe at
\begin{verbatim}http://exep.jpl.nasa.gov/presentations/87-BDT_Pasadena.pdf\end{verbatim}
\bibitem{Crowe}M. J. Crowe. {\it The extraterrestrail Life Debate 1750 - 1900. The Ideas of 
a Plurality of Worlds from Kant to Lowell}. Cambridge University Press (1986)

\bibitem{NASA} FY 2010 NASA Budget Request, 
section Astrophysics p. SCI-189. Available at
\begin{verbatim}
http://www.nasa.gov/pdf/344769main_science_3a_astro.pdf\end{verbatim}
\bibitem{Pathway} 	
	"Pathways Towards Habitable Planets." International Conference. Barcelona (Spain) September 14 - 18 , 2009.
Website available at
\begin{verbatim}
http://www.pathways2009.net/\end{verbatim}
\bibitem{Schneider}J. Schneider. "The question 'Are we alone ?' in different cultures". in 
IAU Symposium 260  {\it The role of Astronomy in Society and Culture}. Cambridge University Press.
Boksemberg and Valls-Gabaud Eds. in press. (2009)

\bibitem{Schneideretal}J. Schneider, A. Leger, M. Fridlund {\it et al.}  "The Far Future of Exoplanet Direct Characterization". {\it Astrobiology}, in press (2009)

\bibitem{Weiler} E. Weiler  NASA Press Release  09-049. Available at 

\begin{verbatim}http://www.nasa.gov/home/hqnews/2009/mar/HQ_09049_Kepler.html
\end{verbatim}
\end{thebibliography}
\end{document}